\documentclass[prd,onecolumn,notitlepage,nofootinbib,longbibliography,superscriptaddress]{revtex4-2}

\usepackage{amsmath,amssymb}
\usepackage[colorlinks=true, allcolors=blue]{hyperref}
\usepackage{graphicx,placeins}
\usepackage[italicdiff]{physics}
\usepackage{comment}
\usepackage{booktabs}
\usepackage{xcolor}
\usepackage{slashed}

\newcommand{\be}{\begin{equation}}
\newcommand{\ee}{\end{equation}}

\makeatletter
\newcommand{\twocolclearpage}{%
  \close@column@grid
  \clearpage
  \twocolumngrid
}
\makeatother

\begin{document}

\title{Electron and photon decays in dimension-six Lorentz-violating QED beyond the asymptotic regime}

\author{J. M. Carmona}
\email{jcarmona@unizar.es}
\affiliation{Departamento de Física Teórica, Universidad de Zaragoza, Zaragoza 50009, Spain}
\affiliation{Centro de Astropartículas y Física de Altas Energías (CAPA), Universidad de Zaragoza}

\author{J. L. Cortés}
\email{cortes@unizar.es}
\affiliation{Departamento de Física Teórica, Universidad de Zaragoza, Zaragoza 50009, Spain}
\affiliation{Centro de Astropartículas y Física de Altas Energías (CAPA), Universidad de Zaragoza}

\author{A. Gkioni}
\email{agkioni@unizar.es}
\affiliation{Departamento de Física Teórica, Universidad de Zaragoza, Zaragoza 50009, Spain}
\affiliation{Centro de Astropartículas y Física de Altas Energías (CAPA), Universidad de Zaragoza}

\author{F. Rescic}
\email{filip.rescic@uniri.hr}
\affiliation{University of Rijeka, Faculty of Physics, Rijeka 51000, Croatia}
\affiliation{Departamento de Física Teórica, Universidad de Zaragoza, Zaragoza 50009, Spain}
\affiliation{Centro de Astropartículas y Física de Altas Energías (CAPA), Universidad de Zaragoza}

\author{M. A. Reyes}
\email{maykoll.reyes@uniri.hr}
\affiliation{University of Rijeka, Faculty of Physics, Rijeka 51000, Croatia}
\affiliation{Centro de Astropartículas y Física de Altas Energías (CAPA), Universidad de Zaragoza}

\author{T. Terzi\'c}
\email{tterzic@phy.uniri.hr}
\affiliation{University of Rijeka, Faculty of Physics, Rijeka 51000, Croatia}

\begin{abstract}
We study the kinematics and dynamics of electron and photon decay processes in a Lorentz-invariance-violating extension of quantum electrodynamics with rotationally invariant higher-dimensional operators of dimension six (quadratic order, $n=2$). As a starting point, we consider the Lagrangian framework introduced in previous works, where Lorentz violation was implemented through additional spatial derivatives. We show that, upon using the Lorentz-invariant equations of motion, this formulation is equivalent to an alternative representation in which the Lorentz-violating terms involve additional temporal derivatives. Working in this latter formulation and employing the standard representation of photon polarization states, we show that decay rates can be computed without introducing modifications of the photon--fermion--fermion vertex. Within this framework, we derive explicit expressions for the decay widths of the processes $e^- \to e^- \gamma$ and $\gamma \to e^- e^+$, going beyond the asymptotic approximation adopted in earlier studies. 
The present framework will be of direct use for subsequent phenomenological analyses of Lorentz-violating effects in high-energy QED processes relevant to the emission, propagation, and detection of very-high-energy gamma rays, including near-threshold regimes.
\end{abstract}

\maketitle
\tableofcontents

\section{Introduction}
Lorentz invariance (LI) is a cornerstone of both the Standard Model and General Relativity.
Nevertheless, many candidate scenarios for physics beyond the Standard Model---and, in particular, several approaches to quantum gravity---suggest that LI may be an emergent symmetry, approximate at low energies and potentially violated at sufficiently high energies. This possibility has motivated a broad experimental and observational program aimed at testing LI with ever-increasing sensitivity~\cite{Addazi:2021xuf}.
Comprehensive reviews of Lorentz-invariance tests and their theoretical frameworks can be found in Refs.~\cite{Mattingly:2005re,Liberati:2009pf,Liberati:2013xla}, while a systematic effective-field-theory (EFT) parametrization of Lorentz and CPT violation is provided by the Standard-Model Extension (SME) framework~\cite{Colladay:1998fq,Kostelecky:2008ts}.

Early systematic treatments of renormalizable (dimension $\le 4$) Lorentz-invariance violation (LIV) in the Standard Model were developed in Refs.~\cite{Coleman:1998ti,Colladay:1998fq}, while LIV encoded in higher-dimension operators was explored in Ref.~\cite{Myers:2003fd}. In the latter case, LIV effects are suppressed by a large ultraviolet scale $\Lambda$ (often associated with the Planck scale $E_\text{Pl}\approx 1.22\times 10^{28}$ eV), a structure that is particularly well motivated from a quantum-gravity perspective. In the QED sector, such operators lead to modified dispersion relations for electrons and photons, opening kinematically new processes and shifting thresholds for reactions that are forbidden in LI kinematics.

Gauge invariance, however, is typically maintained also in Lorentz-violating extensions of QED. For the questions addressed in this work, this requirement is crucial already at tree level. First, it ensures that the photon still propagates only two physical transverse degrees of freedom, so that decay rates can be computed using physical
polarization states without introducing spurious longitudinal contributions.
Second, gauge invariance enforces the appropriate Ward identities, which guarantee that observable quantities such as spin-averaged squared amplitudes and
decay widths are independent of the gauge choice and of the particular representation adopted for the polarization sum. Finally, it strongly constrains the form of the electron--photon interaction, allowing one to derive consistent Feynman rules directly from the
Lagrangian and to relate different operator bases that implement the same boost-breaking, rotationally invariant dynamics---a point that will become explicit when we
compare alternative formulations of the same operator content.

High-energy astrophysics provides an especially sensitive arena for LIV searches when tiny modifications increase with the energy. In the QED sector, the most stringent tests often arise from the (in)stability of particles that are observed to propagate over macroscopic distances at very high energies. In particular, photon decay $\gamma \to e^+e^-$ and vacuum Cherenkov emission $e\to e\gamma$ are typically among the strongest probes of LIV dispersion effects, since the existence of stable VHE/UHE photons and high-energy electrons tightly constrains their modified kinematics.

Threshold analyses and combined constraints for Planck-suppressed LIV were developed early on in Ref.~\cite{Jacobson:2002hd}, and the importance of going beyond threshold kinematics to include actual rates at energies close to the threshold was emphasized in subsequent work~\cite{Jacobson:2005bg}. Indeed, building on theoretical expressions derived for renormalizable models of Lorentz violation, such as those of Refs.~\cite{Klinkhamer:2008ky,Coleman:1997xq}, recent dedicated studies of photon decay and vacuum Cherenkov emission using TeV--PeV gamma-ray data have been used to set stringent limits on photon-sector LIV in the case of $d=4$ operators (see, e.g.,
Refs.~\cite{Martinez-Huerta:2017ulw,HAWC:2019gui}). However, for Planck-suppressed scenarios described by higher-dimension operators and characterized by a large UV scale $\Lambda$, explicit Lagrangian-based calculations of amplitudes and decay rates are
comparatively scarce. To our knowledge, the only systematic treatment of Lorentz-violating spinor electrodynamics that includes both $d=4$ and $d=6$ operators is the work of Rubtsov, Satunin, and Sibiryakov~\cite{Rubtsov:2012kb}, whose results have become standard input in many phenomenological studies. Their formalism implements LIV through spatial-derivative operators in the fermion and photon sectors, yielding compact expressions for several observables in the ultra-relativistic regime where the electron mass can be neglected (``well above threshold''). As a result, the available expressions for decay rates are not directly applicable in the vicinity of the kinematic thresholds, where finite-mass effects control the detailed onset of the process. 

The goal of the present work is to complement and extend these developments by providing a
dynamical framework for electron and photon decays in a dimension-six, CPT-even, rotationally invariant LIV extension of QED that remains valid also in the vicinity of the kinematic thresholds. While asymptotic expressions are often sufficient to derive approximate constraints, a consistent treatment near threshold is required for a controlled description of the opening of the decay channels and for reliable implementations in Monte Carlo simulations and phenomenological predictions.

We adopt a representation in which the fermionic LIV operator involves additional \emph{temporal} derivatives. 
This representation is equivalent to the spatial-derivative implementation considered in Ref.~\cite{Rubtsov:2012kb}, which can be made explicit using the LI equations of motion at the appropriate order in $1/\Lambda^2$. 
However, this choice is technically convenient: it leads to simple external-state constructions (modified spinors and spin sums) while keeping the LI spatial vertex structure, thereby streamlining the calculation of amplitudes. As a non-trivial check, we will verify explicitly that our results reduce to those of Ref.~\cite{Rubtsov:2012kb} in the limit $m_e\to 0$. 

In this work, we concentrate on the processes $\gamma \to e^+e^-$ and $e\to e\gamma$ because their rates are tightly linked to the stability of photons and electrons and therefore yield particularly strong and transparent constraints. 
Moreover, these two channels are governed by the same electron and photon dispersion coefficients and are thus deeply correlated; treating them jointly provides a natural way to assess which classes of LIV QED deformations can be compatible with the simultaneous stability of both species, a point that is obscured when the channels are analyzed in isolation. 
Although these processes have been extensively used to set strong constraints on superluminal dispersion relations for photons and electrons, respectively~\cite{Jacobson:2002hd}, they may also occur in certain subluminal scenarios, depending on the relative modifications of the electron and photon dispersion relations. For instance, a subluminal modification affecting only the photon sector may allow vacuum Cherenkov emission, even though photon decay remains forbidden.
The results presented here will provide the technical foundation for a forthcoming phenomenological study.

The structure of this paper is as follows. In Sec.~\ref{sec:model} we introduce the LIV QED Lagrangian and specify the dimension-six ($n=2$) operator content. We discuss the implementation of gauge invariance, compare alternative representations of the fermionic LIV operators involving spatial or temporal derivatives, and present the corresponding completeness relations for fermion spinors and photon polarization states. In Sec.~\ref{sec:cherenkov} and~\ref{sec:decay}, we analyze vacuum Cherenkov radiation and photon decay, respectively. For each process, we explore the LIV-modified kinematics, compute the corresponding tree-level amplitudes, and perform the phase-space integrations to derive the differential decay widths. Throughout these calculations, we retain the electron mass to accurately identify the kinematic domains in which the channels are open. We conclude in Sec.~\ref{sec:conclusions} with a discussion of the scope of the framework and its relevance for phenomenological applications. 

Throughout this work we use the mostly-minus Minkowski metric
$\eta_{\mu\nu}=\mathrm{diag}(+,-,-,-)$, so that so that $\eta_{ij}=-\delta_{ij}$ and $p_i=\eta_{i\mu}p^\mu=-p^i$.

\section{QED-LIV Lagrangian}
\label{sec:model}

\subsection{Effective Lagrangian and operator content}
\label{sec:model:lagrangian}

We consider a parity- and CPT-even, gauge-invariant extension of QED in which spatial rotations remain exact while boost invariance is broken by higher-dimension operators suppressed by a large UV scale $\Lambda$. 

In the photon sector, Lorentz-violating operators of dimension five are necessarily parity- and CPT-odd and generically induce vacuum birefringence, which is constrained to extremely small levels by astrophysical polarization measurements~
\cite{Myers:2003fd,Kostelecky:2009zp,Kostelecky:2008ts,Liberati:2013xla}. Motivated by this, and following the approach of Ref.~\cite{Rubtsov:2012kb}, we restrict attention to parity- and CPT-even operators also in the fermion sector. 

A generic Lorentz-violating fermionic operator may involve an arbitrary $4\times4$ matrix in Dirac space, which can be expanded on the complete basis
$\{\mathbb{I},\,\gamma^\mu,\,\sigma^{\mu\nu},\,\gamma^5\gamma^\mu,\,\gamma^5\}$.
Parity invariance excludes fermionic operators involving $\gamma^5$. This prevents spin-dependent effects, which is the reason why we will focus on unpolarized decay rates. Under these assumptions, the only CPT-even fermionic operator at dimension five would be proportional to the identity matrix in Dirac space~\cite{Petrov:2025wey}. However, such a term would involve different fermion helicities and would then be suppressed by a factor $m_e/\Lambda$, so that we can exclude it, again following Ref.~\cite{Rubtsov:2012kb}.

As a consequence, the leading fermionic Lorentz-violating contributions arise at dimension six. At this order, CPT-even operators must involve an odd number of gamma matrices, leaving a single rotationally invariant fermionic structure proportional to $\gamma^\mu$. We therefore restrict ourselves to quadratic Lorentz violation, corresponding to dimension-six operators that modify the dispersion relations at
${\cal O}(p^{2+n}/\Lambda^n)$, with $n=2$, and work consistently to leading order in $1/\Lambda^2$.

The Lagrangian is written as
\begin{equation}
\mathcal{L} \;=\; \mathcal{L}_{\rm QED} \;+\; \delta\mathcal{L}^{(6)}_{\gamma}
\;+\; \delta\mathcal{L}^{(6)}_{e}\,,
\end{equation}
with the Lorentz-invariant renormalizable part
\begin{equation}
\mathcal{L}_{\rm QED}
= -\frac14 F_{\mu\nu}F^{\mu\nu}
+ \bar\psi\,(i\slashed{D}-m_e)\,\psi\,,
\qquad
D_\mu = \partial_\mu + i e A_\mu\,.
\end{equation}
Rotational invariance implies that the LIV corrections can only involve spatial contractions, such as $\vec{\gamma}\cdot \vec{\partial}$ and $\partial_i^2\equiv \partial_i\partial_i$, or, alternatively, temporal contractions as $\gamma^0 \partial_0$ and $\partial_0^2\equiv\partial_0 \partial_0$. These two perspectives give rise to two options for parity, CPT-even and gauge-invariant dimension six operators as the minimal set that modify the free dispersion relation of the photon and the electron:\footnote{Up to integration by parts and to field redefinitions that are trivial on-shell at the order considered.}
\begin{align}
\text{(a)}\quad
\delta\mathcal{L}^{(6)}_{\gamma}
&= +\frac{\chi_\gamma}{4\Lambda^2}\,
F_{ij}\,\partial_i^2\,F^{ij}\,,
& \hspace{-2cm}
\delta\mathcal{L}^{(6)}_{e}
&= -\frac{\chi_e}{2\Lambda^2}\,
(iD_j) \bar\psi\,\gamma^i (iD_i) (iD_j) \psi
\label{eq:LIV_spatial}
\\[2mm]
\text{(b)}\quad
\delta\mathcal{L}^{(6)}_{\gamma}
&= -\frac{\chi_\gamma}{2\Lambda^2}\,
F_{0i}\,\partial_0^2\,F^{0i}\,,
& \hspace{-2cm}
\delta\mathcal{L}^{(6)}_{e}
&= -\frac{\chi_e}{2\Lambda^2}\,
\bar\psi\,\gamma^0\,(i D_0)^3\,\psi
\label{eq:LIV_temporal}
\end{align}
where $i,j=1,2,3$ are spatial indices and $\chi_e,\chi_\gamma$ are dimensionless LIV coefficients. The signs and factors accompanying the $\chi_e$, $\chi_\gamma$ coefficients have been chosen so that options (a) and (b) give the same dispersion relations for the photon and the electron. We will refer to option (a) in Eq.~\eqref{eq:LIV_spatial} as the ``spatial-derivative'' representation of LIV in the QED Lagrangian, which is the implementation adopted in Ref.~\cite{Rubtsov:2012kb}, and to option (b) as its ``temporal-derivative'' representation. Note that, since we are treating LIV as a first-order perturbation, introducing high-order temporal derivatives in the Lagrangian does not entail any stability issues in the effective theory.

In Eqs.~\eqref{eq:LIV_spatial}-\eqref{eq:LIV_temporal}, gauge invariance is imposed at the level of the effective Lagrangian by writing the fermionic LIV operator in terms of the covariant derivatives, $D_i$ and $D_0$ respectively, while the photon-sector correction is expressed solely in terms of the field strength tensor. As a result, the theory retains the usual $U(1)$ gauge symmetry of QED: the photon propagates only two physical transverse degrees of freedom, physical observables are independent of the gauge choice, and the structure of interaction vertices is strongly constrained. While the kinematics of the two representations (a) and (b) is trivially the same, since both type of modifications produce the same dispersion relations, the properties implied by gauge invariance will play an important technical role in the derivation of decay amplitudes and in establishing the equivalence, at ${\cal O}(1/\Lambda^2)$, of these two different operator bases.

\subsection{Modified dispersion relations}
\label{sec:model:dispersion}

In the free-field limit ($D_\mu\to\partial_\mu$), varying the spatial-derivative action in Eq.~\eqref{eq:LIV_spatial} yields the modified equations of motion. We adopt the plane-wave ansatz $\psi(x)=\tilde u(\vec p)e^{-ip\cdot x}$ for fermions and $A_\mu(x)=\epsilon_\mu(\vec k)e^{-ik\cdot x}$ for photons (alongside the radiation gauge $\epsilon_0 = 0$, $k_j \epsilon_j = 0$). We explicitly introduce the tilde notation $\tilde u(\vec p)$ because the Lorentz-violating terms deform the Dirac matrix structure, requiring a modified spinor solution. In contrast, the photon wave equation modifies only the scalar dispersion relation, leaving the standard transverse polarizations $\epsilon_i(\vec k)$ unchanged. The resulting equations in momentum space are:
\begin{align}
\Big(\slashed{p}-m_e+\frac{\chi_e}{2\Lambda^2}\,\gamma^i p_i \vec{p}^2\Big)\tilde u(\vec p) &= 0\,, \label{eq:ele_field} \\
\Big(\omega^2 - |\vec{k}|^2 - \frac{\chi_\gamma}{\Lambda^2} |\vec{k}|^4\Big) \epsilon_i(\vec k) &= 0\,. \label{eq:pho_field}
\end{align}

The equivalence of the spatial and temporal bases at $\mathcal{O}(1/\Lambda^2)$ follows directly from the unperturbed equations of motion. For ultrarelativistic electrons, the zeroth-order Dirac equation $\gamma^i p_i\,\tilde u(\vec p)=(-\gamma^0 p_0+m_e)\tilde u(\vec p)$ and the on-shell condition $p_0^2 \simeq \vec p^{\,2}$ allow us to transform the spatial momentum contraction into the temporal one, neglecting subleading mass contributions:
\begin{equation}
\frac{\chi_e}{2\Lambda^2}\,\gamma^i p_i\,\vec p^{\,2}\,\tilde u(\vec p) \simeq -\frac{\chi_e}{2\Lambda^2}\,\gamma^0 p_0^3\,\tilde u(\vec p)\,.
\end{equation}
Similarly, for free photons, the zeroth-order relation $\omega^2 \simeq |\vec{k}|^2$ and the identity $F_{ij}F^{ij} \simeq -2 F_{0i}F^{0i}$ imply that the spatial contraction transforms exactly into the temporal one when evaluated on the plane waves:
\begin{equation}
\frac{\chi_\gamma}{4\Lambda^2}\,F_{ij}\,\partial_k^2\,F^{ij} \simeq -\frac{\chi_\gamma}{2\Lambda^2}\,F_{0i}\,\partial_0^2\,F^{0i}\,.
\end{equation}
This demonstrates that the spatial and temporal-derivative operators in Eqs.~\eqref{eq:LIV_spatial} and~\eqref{eq:LIV_temporal}, respectively, reproduce the same leading-order corrections, ensuring both formulations describe identical physics.

Solving the modified free-field equations \eqref{eq:ele_field} and \eqref{eq:pho_field} yields the dispersion relations at leading order in $1/\Lambda^2$:
\begin{align}
\omega^2 &= |\vec{k}|^2 \Big(1 + \chi_\gamma\,\frac{|\vec{k}|^2}{\Lambda^2}\Big)\,, \label{eq:disp_photon} \\ 
E^2 &= |\vec{p}|^2 + m_e^2 + \chi_e\,\frac{|\vec{p}|^4}{\Lambda^2}\,, \label{eq:disp_electron} 
\end{align}
which expand in the ultrarelativistic limit to:
\begin{align} 
\omega &\simeq |\vec{k}| + \chi_\gamma\,\frac{|\vec{k}|^3}{2\Lambda^2}\,, \label{eq:omega_UR} \\ 
E &\simeq |\vec{p}| + \frac{m_e^2}{2|\vec{p}|} + \chi_e\,\frac{|\vec{p}|^3}{2\Lambda^2}\,. \label{eq:E_UR} 
\end{align}

From the previous equations, one can obtain the modified group velocities for the photon and electron, respectively,
\begin{align}
    v_\gamma\equiv \frac{d\omega}{d|\vec k|}&\simeq 1+\frac{3}{2}\chi_\gamma \frac{|\vec k|^2}{\Lambda^2}\,, \\
    v_e \equiv \frac{dE}{d|\vec p|}&\simeq 1-\frac{1}{2}\frac{m_e^2}{|\vec p|^2}+\frac{3}{2}\chi_e\frac{|\vec p|^2}{\Lambda^2}\,.
\end{align}
Dispersion relations with positive (negative) values of the Lorentz-violating coefficients $\chi_\gamma$ and $\chi_e$ are commonly referred to as superluminal (subluminal), since the corresponding group velocities satisfy $v>1$ ($v<1$) in the high-energy regime where the LIV corrections dominate over the mass term. While superluminal dispersion relations are usually related with decay processes, we will see that they can happen even with subluminal dispersion relations, something which has not been sufficiently stressed in the literature.

\subsection{Modified Feynman rules}
\label{sec:model:feynman}

\subsubsection{Spatial representation}

In the spatial representation given by Eq.~\eqref{eq:LIV_spatial}, the modified free fermion equation of motion can be written as a standard Dirac equation involving a deformed momentum $\tilde p^\mu$ whose spatial components are shifted,
\begin{equation}
(\slashed{\tilde p}-m_e)\,\tilde u(\vec p)=0\,, \qquad \tilde p_0=p_0\,,\qquad \tilde p_i = p_i\Big(1+\chi_e\frac{\vec p^{\,2}}{2\Lambda^2}\Big)\,,
\label{eq:tildep_spatial}
\end{equation}
where $p_0$ is the energy $E$ appearing in the dispersion relation, Eq.~\eqref{eq:disp_electron}. In the following we will see that enforcing gauge invariance of the one-photon emission amplitude requires a corresponding modification of the fermion--fermion--photon vertex, since $k_\mu \bar{\tilde u}(\vec p\,')\gamma^\mu \tilde u(\vec p)$ no longer vanishes by the usual Lorentz-invariant Dirac equation argument.

Consider the one-photon emission amplitude (vacuum Cherenkov process) $e^-(p)\to e^-(p')+\gamma(k)$, which in Lorentz-invariant QED reads $\mathcal{A}=\epsilon_\mu^*(\vec k)\,\mathcal{M}^\mu$,
where $e$ is the electron charge and $\mathcal{M}^\mu\equiv \bar{\tilde u}(\vec p\,')\,(-ie)\gamma^\mu\,\tilde u(\vec p)$. Gauge invariance implies the Ward identity $k_\mu\mathcal{M}^\mu=0$, i.e.
\begin{equation}
k_\mu\,\bar{\tilde u}(\vec p\,')\,(-ie)\gamma^\mu\,\tilde u(\vec p)=0\,,
\label{eq:ward_basic}
\end{equation}
which in Lorentz-invariant QED follows from $k_\mu=p_\mu-p'_\mu$ together with the Dirac equations $\slashed{p}\,u(\vec p)=m_e u(\vec p)$ and $\bar u(\vec p\,') \slashed{p}'=m_e \bar u(\vec p\,')$. In the Lorentz-violating case, however, the external spinors satisfy $\slashed{\tilde p}\,\tilde u(\vec p)=m_e \tilde u(\vec p)$ rather than $\slashed{p}\,u(\vec p)=m_e u(\vec p)$, and one finds
\begin{align}
k_\mu\,\bar{\tilde u}(\vec p\,')\,(-ie)\gamma^\mu\,\tilde u(\vec p) &= (-ie)\,\bar{\tilde u}(\vec p\,')(\slashed{p}-\slashed{p}')\tilde u(\vec p) \nonumber\\
&= (-ie)\,\bar{\tilde u}(\vec p\,')\big[(\slashed{p}-\slashed{\tilde p})-(\slashed{p}'-\slashed{\tilde p}')\big]\tilde u(\vec p) \nonumber\\
&= \frac{ie\,\chi_e}{2\Lambda^2}\sum_i\big(p_i\,\vec p^{\,2}-p'_i\,\vec p\,'^{\,2}\big)\,\bar{\tilde u}(\vec p\,')\gamma^i \tilde u(\vec p)\,,
\label{eq:ward_spatial_nonzero}
\end{align}
which is generically non-vanishing. Gauge invariance therefore requires a modified fermion--fermion--photon vertex,
\begin{equation}
\mathcal{A}_{\rm LIV}=\epsilon_\mu^*(k)\,\bar{\tilde u}(\vec p\,')\,(-ie)\Gamma^\mu\,\tilde u(\vec p)\,, \qquad \Gamma^\mu=\gamma^\mu+\sum_i\gamma^i\,V_i^{\mu}\,,
\label{eq:Gamma_spatial_def}
\end{equation}
with $V_i^\mu$ fixed by imposing $k_\mu\bar{\tilde u}(\vec p\,')\Gamma^\mu \tilde u(\vec p)=0$. This yields the condition
\begin{equation}
(p_\mu-p'_\mu)\,V_i^\mu=\frac{\chi_e}{2\Lambda^2}\big(p_i\,\vec p^{\,2}-p'_i\,\vec p\,'^{\,2}\big)\,.
\label{eq:Vconstraint_spatial}
\end{equation}
A convenient mass-independent choice is $V_i^0=0$, which leads to
\begin{equation}
V_i^k=\frac{\chi_e}{2\Lambda^2}\big(p_i p^k+p'_i p'^k+\delta_i^k\,\vec p\cdot\vec p\,'\big)\,.
\label{eq:Vsolution_spatial}
\end{equation}
This result agrees with the photon--electron--positron vertex deformation in Eq.~(8) of Ref.~\cite{Rubtsov:2012kb}, which is obtained from Eq.~\eqref{eq:Vsolution_spatial} for the electron--photon--electron vertex with the replacement $\vec p\,'\mapsto -\vec p\,'$. The modification of the photon-electron-electron vertex, Eq.~\eqref{eq:Vsolution_spatial}, can be derived directly by considering, in the fermionic operator in Eq.~\eqref{eq:LIV_spatial}, the contribution proportional to the gauge field in one of the three covariant derivatives.

\subsubsection{Temporal representation}

A similar trick can be made in the temporal representation given by Eq.~\eqref{eq:LIV_temporal}. Here the deformed momentum entering the generalized Dirac equation instead has a modified time component,
\begin{equation}
\slashed{\tilde p}\,\tilde u(\vec p)=m_e\,\tilde u(\vec p)\,, \qquad \tilde p_0=p_0\Big(1-\chi_e\frac{p_0^2}{2\Lambda^2}\Big)\,,\qquad \tilde p_i=p_i\,.
\label{eq:tildep_temporal}
\end{equation}
In this basis, the Lorentz-invariant Ward identity \eqref{eq:ward_basic} is again violated if one uses the bare vertex $(-ie)\gamma^\mu$ with external spinors obeying Eq.~\eqref{eq:tildep_temporal}. Proceeding as above, one finds
\begin{equation}
k_\mu\,\bar{\tilde u}(\vec p\,')\,(-ie)\gamma^\mu\,\tilde u(\vec p)=-\frac{ie\,\chi_e}{2\Lambda^2}\big(p_0^3-p_0'^3\big)\,\bar{\tilde u}(\vec p\,')\gamma^0 \tilde u(\vec p)\,,
\label{eq:ward_temporal_nonzero}
\end{equation}
so gauge invariance requires a modified vertex of the form
\begin{equation}
\Gamma^\mu=\gamma^\mu+\gamma^0 V_0^\mu\,, \qquad (p_\mu-p'_\mu)\,V_0^\mu=-\frac{\chi_e}{2\Lambda^2}\big(p_0^3-p_0'^3\big)\,.
\label{eq:Gamma_temporal_def}
\end{equation}
If one demands a mass-independent choice, one has
\begin{equation}
V_0^i=0\,, \qquad V_0^0=-\frac{\chi_e}{2\Lambda^2}\big(p_0^2+p_0'^2+p_0 p_0'\big)\,.
\label{eq:Vsolution_temporal}
\end{equation}

A crucial technical advantage emerges in this temporal-derivative representation: the spatial components of the vertex remain entirely equal to their Lorentz-invariant form, $\Gamma^i=\gamma^i$. For this reason, we adopt the temporal basis for the remainder of this work.

Because the spatial vertex is undeformed, the photon polarization sums proceed exactly as in the standard theory. As a gauge theory, QED contains massless spin-one photons with two transverse polarization states. We therefore work directly with physical polarization vectors $\epsilon_\mu^{(\lambda)}(\vec k)$ satisfying $k^\mu \epsilon_\mu^{(\lambda)}(\vec k)=0$. By choosing the Coulomb gauge ($\epsilon_0^{(\lambda)}=0$), the photon lacks a temporal polarization component, ensuring that the LIV vertex correction $V_0^0$ never contributes to physical amplitudes. In computations of spin- or polarization-averaged squared amplitudes, it is thus convenient to use the standard transverse projector in three-dimensional form:
\begin{equation}
\sum_{\lambda=1,2}\epsilon_i^{(\lambda)}(\vec k)\,\epsilon_j^{(\lambda)}(\vec k)^{*} = \delta_{ij}-\frac{k_i k_j}{\vec k^{\,2}}\,.
\label{eq:transverse_projector}
\end{equation}

Finally, to utilize standard trace techniques alongside these photon projectors, the external fermion states must be properly normalized. From the dispersion relation \eqref{eq:disp_electron}, the deformed temporal component evaluates exactly to
\begin{equation}
    \tilde p_0=\sqrt{\vec{p}^2+m_e^2}\,.
    \label{eq:tilde-disp-rel}
\end{equation}
In other words, the deformed four-momentum $\tilde p^\mu= (\tilde p_0,\vec p)$ satisfies the standard special relativistic dispersion relation. Therefore, by using this deformed energy to normalize the spinors, the completeness relations retain the exact formal structure of Lorentz-invariant QED:
\begin{equation}
\sum_s \tilde u^{(s)}(\vec p)\,\bar{\tilde u}^{(s)}(\vec p) = \slashed{\tilde p}+m_e\,, \qquad \sum_s \tilde v^{(s)}(\vec p)\,\bar{\tilde v}^{(s)}(\vec p) = \slashed{\tilde p}-m_e\,.
\label{eq:spin_sums}
\end{equation}

Consequently, amplitude computations parallel those in the standard formulation. Spinor traces can be evaluated using standard algebraic rules, with all Lorentz-violating effects safely and consistently captured entirely through the modified kinematics encoded in $\tilde p^\mu$.

\section{Vacuum Cherenkov radiation}
\label{sec:cherenkov}

We first consider vacuum Cherenkov radiation,
\begin{equation}
e^-(p)\to e^-(p')+\gamma(k),
\end{equation}
which is forbidden in Special Relativity (SR) but may become kinematically allowed in LIV scenarios. In this section, we analyze the kinematics and dynamics of this process.
We will repeat the same procedure for the photon decay in Section~\ref{sec:decay}.

\subsection{Kinematics}
\label{sec:cherenkov:kinematics}

We consider the decay of an incoming electron with three-momentum $\vec p$ and energy $E$ into a final-state electron and photon. It is convenient to parameterize the final-state phase space in terms of a longitudinal momentum fraction $x\in [0,1]$ carried by the photon and a transverse momentum $\vec k_T$ orthogonal to $\vec p$,
\begin{equation}
\vec p\,' = (1-x)\,\vec p - \vec k_T\,, \qquad
\vec k  = x\,\vec p + \vec k_T\,.
\label{eq:momentum-decomposition}
\end{equation}
With this parameterization, momentum conservation is automatic, while energy conservation fixes $\vec k_T^{\,2}$ as a function of $x$.

Since we are interested in vacuum Cherenkov emission at high energies, we will use the ultrarelativistic expansions of the modified dispersion relations, Eqs.~\eqref{eq:omega_UR} and \eqref{eq:E_UR}, along with the approximation in which the transverse momentum satisfies $\vec k_T^{\,2} \ll |\vec p|^2$, as in Ref.~\cite{Rubtsov:2012kb}. Unlike in that work, however, we retain the leading electron-mass contributions, which become important near threshold. We will refer to this as the `collinear approximation', meaning that the dominant term in the energy of each particle is given by the modulus of the longitudinal momentum along the direction of the decaying particle. While this is obviously the case of the incoming electron, which is by assumption ultrarelativistic, one might wonder about the validity of this approximation for the final particles; for example, the outgoing electron could in principle become non-relativistic if most of the momentum were transferred to the photon. However, we show in Appendix~\ref{app:URvalidity} that the computation of the decay width within the collinear approximation is self-consistent: the region where this approximation could break down gives only subleading contributions, suppressed by powers of $|\vec p|^2/\Lambda^2$. As a result, we can approximate the energies of the particles by
\begin{align}
E &\simeq |\vec p| + \frac{m_e^2}{2|\vec p|} + \chi_e\frac{|\vec p|^3}{2\Lambda^2}\,, \label{eq:VC-E-UR}\\
E' &\simeq (1-x)|\vec p| + \frac{m_e^2+\vec k_T^{\,2}}{2(1-x)|\vec p|} + \chi_e\frac{(1-x)^3|\vec p|^3}{2\Lambda^2}\,, \label{eq:VC-E'-UR}\\
\omega &\simeq x|\vec p| + \frac{\vec k_T^{\,2}}{2x|\vec p|} + \chi_\gamma\frac{x^3|\vec p|^3}{2\Lambda^2}\,. \label{eq:VC-omega-UR}
\end{align}
Imposing energy conservation, $E=E'+\omega$, one finds
\begin{equation}
\vec k_T^{\,2}=x^2\left[\frac{|\vec p|^4}{\Lambda^2}\big(x^2(1-x)(\chi_e-\chi_\gamma)+3(1-x)^2\chi_e\big)-m_e^2\right] .
\label{eq:kT2_VC}
\end{equation}
The decay is kinematically allowed only if $\vec k_T^{\,2}(x)\ge 0$ for some $x\in[0,1]$, which determines the allowed interval of the longitudinal fraction $x$,
\begin{equation}
(\chi_e-\chi_\gamma)\,x^2(1-x)+3\chi_e(1-x)^2 \;\ge\;\tau_e\,,
\qquad \text{where} \quad
\tau_e\equiv \frac{m_e^2\Lambda^2}{E^4}\,.
\label{eq:VC_condition_tau}
\end{equation}
Depending on the signs and relative size of $(\chi_e,\chi_\gamma)$, this inequality may have no solution (stable electron) or define an interval $[x_{\min},x_{\max}]$ (allowed vacuum Cherenkov emission). In particular, if $\chi_e\le0$ and $\chi_e\le\chi_\gamma$, the left-hand side of Eq.~\eqref{eq:VC_condition_tau} is negative for all $x\in[0,1]$ and no decay is possible. 

In the asymptotic regime $E^4/(m_e^2\Lambda^2)\gg 1$ (i.e., $\tau_e\ll1$), the allowed values of $x$ are determined by
\begin{equation}
(1-x)\left[
    (\chi_e-\chi_\gamma)\,x^2+3\chi_e(1-x)
\right]\ge 0\,.
\end{equation}
Since $x\in[0,1]$, the factor $(1-x)$ is non-negative and the kinematic domain is controlled by the sign of the quadratic polynomial in brackets. When $\chi_e\ge 0$ and $\chi_\gamma\le \chi_e$, this polynomial is non-negative over the entire interval $x\in[0,1]$, and one recovers the ``well above threshold'' result of Ref.~\cite{Rubtsov:2012kb}. For $\tau_e\neq 0$, however, the condition $\vec k_T^{\,2}(x)\ge0$ introduces an additional negative contribution that shortens the allowed range of $x$. A detailed discussion of the kinematic domain as a function of $\chi_e$ and $\chi_\gamma$ will be presented in a forthcoming work.

\subsection{Amplitude and squared matrix element}
\label{sec:cherenkov:amplitude}

The tree-level amplitude for vacuum Cherenkov emission in the temporal-derivative representation, Eq.~\eqref{eq:LIV_temporal}, is
\begin{equation}
\mathcal{A}
=-ie\,\epsilon_\mu^*(\vec k)\,
\bar{\tilde u}(\vec p\,')\,\Gamma^\mu\,\tilde u(\vec p)
=-ie\,\epsilon_i^*(\vec k)\,
\bar{\tilde u}(\vec p\,')\,\gamma^i\,\tilde u(\vec p)\,,
\label{eq:VC_amplitude}
\end{equation}
where we have used that $\epsilon_0(\vec k)=0$, so that $\epsilon_i(\vec k)$ denote the physical transverse photon polarizations, and the fact that in the temporal-derivative representation the spatial components of the fermion--fermion--photon vertex coincide with their Lorentz-invariant form, $\Gamma^i=\gamma^i$ (Eqs.~\eqref{eq:Gamma_temporal_def} and~\eqref{eq:Vsolution_temporal}). The spinors $\tilde u$ satisfy a standard Dirac equation with deformed momentum $\tilde p^\mu$ (Eq.~\eqref{eq:tildep_temporal}), whose components satisfy a standard, Lorentz-invariant, dispersion relation (Eq.~\eqref{eq:tilde-disp-rel}). In this way, the computation of the dynamics of the process will be the same as if we were in SR.

The squared amplitude, averaged over the initial electron spin and summed over final spins and photon polarizations, is given by
\begin{equation}
\overline{|\mathcal A|^2}
=\frac{1}{2}\sum_{\text{s},\,\lambda}|\mathcal A|^2
=\frac{e^2}{2}
\sum_{\lambda}
\epsilon_i^{(\lambda)}(\vec k)\,
\epsilon_j^{(\lambda)}(\vec k)^*\,
\Tr\!\left[
\gamma^i(\slashed{\tilde p}+m_e)\,
\gamma^j(\slashed{\tilde p}'+m_e)
\right],
\label{eq:VC_A2_def}
\end{equation}
where the factor $1/2$ accounts for the average over the initial electron spin. Using the completeness relations \eqref{eq:spin_sums} and the standard transverse polarization sum, \eqref{eq:transverse_projector}, one obtains
\begin{equation}
\overline{|\mathcal A|^2}
=\frac{e^2}{2}
\left(\delta_{ij}-\frac{k_i k_j}{\vec k^{\,2}}\right)
\Tr\!\left[
\gamma^i(\slashed{\tilde p}+m_e)\,
\gamma^j(\slashed{\tilde p}'+m_e)
\right].
\label{eq:VC_A2_trace}
\end{equation}

The fermionic trace can be evaluated using standard Dirac algebra, yielding
\begin{equation}
\Tr\!\left[
\gamma^i(\slashed{\tilde p}+m_e)\,
\gamma^j(\slashed{\tilde p}'+m_e)
\right]
=
4\left(
p^i p'^j + p'^i p^j
+\delta^{ij}\,(\tilde p\!\cdot\!\tilde p'-m_e^2)
\right).
\label{eq:VC_trace}
\end{equation}
Substituting this result into Eq.~\eqref{eq:VC_A2_trace}, the spin-averaged squared amplitude can be written as
\begin{equation}
\overline{|\mathcal A|^2}
=
4e^2\left[
(\vec p\!\cdot\!\vec p\,')
-\frac{(\vec k\!\cdot\!\vec p)(\vec k\!\cdot\!\vec p\,')}{\vec k^{\,2}}
+(\tilde p\!\cdot\!\tilde p')-m_e^2
\right].
\label{eq:VC_A2_compact}
\end{equation}
From Eq.~\eqref{eq:momentum-decomposition}, the scalar products appearing in Eq.~\eqref{eq:VC_A2_compact} can be expressed in terms of the longitudinal momentum fraction $x$ and the transverse momentum $\vec k_T$. In particular, one finds
\begin{align}
(\vec p\!\cdot\!\vec p\,')+(\tilde p\!\cdot\!\tilde p')&= \tilde p_0\,\tilde p'_0
\simeq (1-x)\,\vec p^{\,2}
+\frac{\vec k_T^{\,2}+[1+(1-x)^2]\,m_e^2}{2(1-x)}\,,
\label{eq:VC_scalar1}
\\
\frac{(\vec k\!\cdot\!\vec p)(\vec k\!\cdot\!\vec p\,')}{\vec k^{\,2}}
&\simeq (1-x)\,\vec p^{\,2}-\frac{\vec k_T^{\,2}}{x^2}\,,
\label{eq:VC_scalar2}
\end{align}
where we have used the ultrarelativistic expansions of the special-relativistic dispersion relation Eq.~\eqref{eq:tilde-disp-rel} and retained the leading mass-dependent terms.

Substituting these expressions into Eq.~\eqref{eq:VC_A2_compact}, the spin-averaged squared amplitude can be written as
\begin{equation}
\overline{|\mathcal A|^2}
\simeq
\frac{2e^2\,[1+(1-x)^2]}{(1-x)}\left(\frac{\vec k_T^{\,2}}{x^2}+m_e^2\right)-4e^2m_e^2\,.
\label{eq:VC_A2_kT}
\end{equation}

Thanks to the temporal representation, we have been able to follow a derivation formally identical to one in SR. All the LIV corrections are encoded in the dependence of $\vec k_T^{\,2}$ on $\Lambda$ that comes from the kinematics of the process, Eq.~\eqref{eq:kT2_VC}.

\subsection{Differential decay width}
\label{sec:cherenkov:decay_width}

The decay width for vacuum Cherenkov emission, $e^-(p)\to e^-(p')\gamma(k)$, of an electron with four-momentum $p^\mu\equiv (E,\vec{p})$ is
\begin{equation}
\Gamma_{e\to e\gamma}
=\frac{1}{2E}\int
\frac{d^3\vec p\,'}{(2\pi)^3\,2E'}\,
\frac{d^3\vec k}{(2\pi)^3\,2\omega}\,
(2\pi)^4\,
\delta^{(3)}(\vec p-\vec p\,'-\vec k)\,
\delta(E-E'-\omega)\,
\overline{|\mathcal A|^2}\,,
\label{eq:Gamma_VC_start}
\end{equation}
where $\omega(\vec{k})$ and $E'(\vec p\,')$ are the energies of the outgoing photon and electron, and $\overline{|\mathcal A|^2}$ is the spin-averaged squared amplitude computed in Sec.~\ref{sec:cherenkov:amplitude}. Using the three-momentum delta function to perform the $\vec p\,'$ integral, we get
\begin{equation}
\Gamma_{e\to e\gamma}
=\frac{1}{2E}\int
\frac{d^3\vec k}{(2\pi)^2}\,
\frac{1}{4\,E'\,\omega}\,
\delta\big(E-E'-\omega\big)\,
\overline{|\mathcal A|^2}\,,
\label{eq:Gamma_VC_after_delta3}
\end{equation}
where $E'$ is now a function of $(\vec p-\vec k)$.

Using the parametrization given by Eq.~\eqref{eq:momentum-decomposition}, we have
\begin{equation}
\vec k = x|\vec p| \,\hat p +\vec k_T\,,\qquad
\vec k_T\cdot\vec p=0\,,
\label{eq:VC_kparallel_ktransverse}
\end{equation}
and the measure becomes
\begin{equation}
d^3\vec k
= |\vec p|\,dx\,d^2\vec k_T
= |\vec p|\,dx \,k_T\,dk_T\,d\varphi
= \frac{|\vec p|}{2}\,dx\,d(\vec k_T^{\,2})\,d\varphi\,,
\label{eq:measure_k_to_x_kT}
\end{equation}
where we have used polar coordinates $(k_T,\varphi)$ in the transverse plane to the direction $\hat{p}$. Since the remaining integrand depends on $\vec k_T$ only through $\vec k_T^{\,2}$, the azimuthal integral gives a factor $2\pi$.

The energies $\omega$ and $E'$ depend now on $x$ and $\vec k_T^{\,2}$. Defining
\begin{equation}
F_{\rm VC}(x,\vec k_T^{\,2})\equiv E - E'(x,\vec k_T^{\,2})-\omega(x,\vec k_T^{\,2})\,,
\end{equation}
the energy delta function can be used to integrate over $\vec k_T^{\,2}$:
\begin{equation}
\delta\big(F_{\rm VC}(x,\vec k_T^{\,2})\big)
=\sum_{a}\frac{\delta\big(\vec k_T^{\,2}-\vec k_{T,a}^{\,2}(x)\big)}
{\left|\partial F_{\rm VC}/\partial \vec k_T^{\,2}\right|_{\vec k_T^{\,2}=\vec k_{T,a}^{\,2}(x)}}\,,
\label{eq:delta_F_general}
\end{equation}
where $\vec k_{T,a}^{\,2}(x)$ are the (non-negative) roots of $F_{\rm VC}=0$.
Within the collinear approximation introduced above, $F_{\rm VC}$ is linear in $\vec k_T^{\,2}$ and there is a single root, which is precisely Eq.~\eqref{eq:kT2_VC}:
\begin{equation}
\vec k_T^{\,2}(x)=x^2\left[\frac{|\vec p|^4}{\Lambda^2}\big(x^2(1-x)(\chi_e-\chi_\gamma)
+3(1-x)^2\chi_e\big)-m_e^2\right].
\tag{\ref{eq:kT2_VC}}
\end{equation}
Using Eqs.~\eqref{eq:VC-E'-UR} and~\eqref{eq:VC-omega-UR}, we readily obtain
\begin{equation}
\frac{\partial F_{\rm VC}}{\partial \vec k_T^{\,2}}
= -\frac{1}{2|\vec p|}\left(\frac{1}{1-x}+\frac{1}{x}\right)
= -\frac{1}{2|\vec p|\,x(1-x)}\,,
\label{eq:dFdkT2_VC}
\end{equation}
so that
\begin{equation}
\int d(\vec k_T^{\,2})\,
\delta\big(F_{\rm VC}(x,\vec k_T^{\,2})\big)\,G(x,\vec k_T^{\,2})
=2|\vec p|\,x(1-x)\,G\big(x,\vec k_T^{\,2}(x)\big)\,.
\label{eq:delta_integral_VC}
\end{equation}
Putting everything together, Eq.~\eqref{eq:Gamma_VC_after_delta3} yields
\begin{equation}
\Gamma_{e\to e\gamma}
=\frac{1}{16\pi\,E}
\int dx\,
\frac{|\vec p|^2\,x(1-x)}{E'\,\omega}\,
\overline{|\mathcal A|^2}\Big|_{\vec k_T^{\,2}=\vec k_T^{\,2}(x)}\,,
\label{eq:Gamma_VC_master}
\end{equation}
where $E'=E'(x,\vec k_T^{\,2}(x))$ and $\omega=\omega(x,\vec k_T^{\,2}(x))$.
In the ultrarelativistic limit one may further set $|\vec p|\simeq E$,
$E'\simeq (1-x)E$ and $\omega\simeq xE$, so that the prefactor simplifies to
$|\vec p|^2 x(1-x)/(E'\omega)\simeq 1$ and
\begin{equation}
\Gamma_{e\to e\gamma}
\simeq \frac{1}{16\pi\,E}\int dx\,
\overline{|\mathcal A|^2}\Big|_{\vec k_T^{\,2}=\vec k_T^{\,2}(x)}\equiv \int dx\;\frac{d\Gamma_{e\to e\gamma}}{dx}\,,
\label{eq:Gamma_VC_simplified_prefactor}
\end{equation}
where $d\Gamma_{e\to e\gamma}/dx$ is the differential decay width.

Using $\vec k_T^{\,2}(x)$ from Eq.~\eqref{eq:kT2_VC}, the combination that appears in the squared amplitude Eq.~\eqref{eq:VC_A2_kT} simplifies to
\begin{equation}
\frac{\vec k_T^{\,2}(x)}{x^2}+m_e^2
=
\frac{|\vec p|^4}{\Lambda^2}\Big[x^2(1-x)(\chi_e-\chi_\gamma)+3(1-x)^2\chi_e\Big]
=
(1-x)\frac{|\vec p|^4}{\Lambda^2}\Big[x^2(\chi_e-\chi_\gamma)+3(1-x)\chi_e\Big].
\label{eq:VC_kT_combo}
\end{equation}
Therefore the spin-averaged squared amplitude becomes
\begin{equation}
\overline{|\mathcal A|^2}\Big|_{\vec k_T^{\,2}=\vec k_T^{\,2}(x)}
\simeq
2e^2\,[1+(1-x)^2]\,
\Big[x^2(\chi_e-\chi_\gamma)+3(1-x)\chi_e\Big]
\frac{|\vec p|^4}{\Lambda^2}
\;-\;4e^2 m_e^2\,.
\label{eq:VC_A2_on_shell}
\end{equation}
Using $\alpha\equiv e^2/(4\pi)$, the differential decay width becomes
\begin{equation}
\frac{d\Gamma_{e\to e\gamma}}{dx}
\simeq
\frac{\alpha\,m_e^2}{E} 
\left\{
[1+(1-x)^2]\,
\Big[x^2(\chi_e-\chi_\gamma)+3(1-x)\chi_e\Big]
\frac{E^4}{2m_e^2\Lambda^2}
\;-\;1
\right\},
\label{eq:Gamma_VC_finalform}
\end{equation}
where we used $|\vec p|\simeq E$ at the order considered.

\subsection{Comparison with asymptotic limit}
\label{sec:cherenkov:rubtsov}

Rubtsov et al.~\cite{Rubtsov:2012kb} obtained the following result for the differential decay width of vacuum Cherenkov radiation in the ultrarelativistic regime and well above threshold:
\begin{equation}
\frac{d\Gamma^\text{asymp}_{e\to e\gamma}}{dx}
=
\alpha\left(\frac{2}{x}-2+x\right)\,
\omega'_{\rm LV}(x)\,,
\label{eq:Rubtsov_dGdx}
\end{equation}
where $\omega'_{\rm LV}(x)$ encodes the Lorentz-violating energy mismatch induced by the modified dispersion relations. In the notation of Ref.~\cite{Rubtsov:2012kb}, and restricting to dimension-six operators, it reads
\begin{equation}
\omega'_{\rm LV}(x)
=
x\,p\left[
\frac{g\,p^2}{M^2}(3-3x+x^2)
-\frac{\xi\,p^2}{2M^2}x^2
\right],
\label{eq:Rubtsov_omegaLV}
\end{equation}
where $p$ is the momentum of the incoming electron and $(g,\xi)$ are the dimension-six LIV coefficients in their convention. Using the parameter identification
\begin{equation}
\frac{2g}{M^2}\;\leftrightarrow\;\frac{\chi_e}{\Lambda^2}\,,
\qquad
\frac{\xi}{M^2}\;\leftrightarrow\;\frac{\chi_\gamma}{\Lambda^2}\,,
\label{eq:Rubtsov_identification}
\end{equation}
and writing $p\simeq E$, Eq.~\eqref{eq:Rubtsov_omegaLV} becomes
\begin{equation}
\omega'_{\rm LV}(x)
=
x\,E\,
\frac{E^2}{\Lambda^2}
\left[
\frac{\chi_e}{2}\,(3-3x+x^2)
-\frac{\chi_\gamma}{2}\,x^2
\right].
\label{eq:omegaLV_ours}
\end{equation}

Substituting this expression into Eq.~\eqref{eq:Rubtsov_dGdx} and rearranging the
polynomial factors of $x$, one finds
\begin{equation}
\frac{d\Gamma^\text{asymp}_{e\to e\gamma}}{dx}
=
\frac{\alpha E^3}{2\Lambda^2}
[1+(1-x)^2]\,
\Big[x^2(\chi_e-\chi_\gamma)+3(1-x)\chi_e\Big],
\label{eq:Rubtsov_dGdx_rewritten}
\end{equation}
which coincides exactly with the leading Lorentz-violating contribution to our result in Eq.~\eqref{eq:Gamma_VC_finalform} in the limit $m_e\to0$.

The additional constant term $-\alpha m_e^2/E$ appearing in Eq.~\eqref{eq:Gamma_VC_finalform} represents the leading mass-dependent correction, which is consistently neglected in Ref.~\cite{Rubtsov:2012kb}, where the analysis is restricted to the asymptotic regime well above threshold. Our expression thus extends their result by retaining the full electron-mass dependence that controls the onset of the process near threshold, while reproducing their formula in the appropriate high-energy limit.

\section{Photon decay}
\label{sec:decay}

\subsection{Kinematics}

We consider the decay of a photon with three-momentum $\vec k$ and energy $\omega$ into an electron--positron pair, i.e., $\gamma(k)\to e^-(p_-)+e^+(p_+)$. The final-state momenta are parameterized in terms of a longitudinal momentum fraction $x$ carried by the electron and a transverse momentum $\vec p_T$ orthogonal to $\vec k$,
\begin{equation}
\vec p_- = x\,\vec k + \vec p_T\,, \qquad
\vec p_+ = (1-x)\,\vec k - \vec p_T\,,
\label{eq:momentum-decomposition-PP}
\end{equation}
with $\vec p_T\!\cdot\!\vec k=0$. As before, momentum conservation is automatically satisfied, and energy conservation fixes $\vec p_T^{\,2}$ as a function of $x$.

The energies of the particles in the collinear approximation are
\begin{align}
\omega &\simeq |\vec k| + \chi_\gamma\frac{|\vec k|^3}{2\Lambda^2}\,,
\\
E_- &\simeq x|\vec k| + \frac{m_e^2+\vec p_T^{\,2}}{2x|\vec k|}
+ \chi_e\frac{x^3|\vec k|^3}{2\Lambda^2}\,,
\\
E_+ &\simeq (1-x)|\vec k|
+ \frac{m_e^2+\vec p_T^{\,2}}{2(1-x)|\vec k|}
+ \chi_e\frac{(1-x)^3|\vec k|^3}{2\Lambda^2}\,.
\end{align}

Imposing energy conservation, $\omega=E_-+E_+$, one finds
\begin{equation}
\vec p_T^{\,2}
= x(1-x)\frac{|\vec k|^4}{\Lambda^2}
\Big[\chi_\gamma-(x^3+(1-x)^3)\chi_e\Big]
- m_e^2 \,.
\label{eq:pT2_PP}
\end{equation}
The decay is kinematically allowed only if $\vec p_T^{\,2}\ge0$, which determines the allowed interval of the longitudinal momentum fraction $x$. 

Rearranging the polynomial in brackets, this condition reads
\begin{equation}
(\chi_\gamma-\chi_e)\,x(1-x)
+3\chi_e\,x^2(1-x)^2
\;\ge\;
\tau_\gamma\,,
\qquad \text{where} \quad
\tau_\gamma\equiv\frac{m_e^2\Lambda^2}{\omega^4}\,,
\label{eq:PP_condition_tau}
\end{equation}
with $x\in[0,1]$.

Introducing the symmetric variable
\begin{equation}
u\equiv x(1-x)\,,\qquad 0\le u\le\frac14\,,
\label{eq:u_def_PP}
\end{equation}
the inequality becomes
\begin{equation}
f(u)\equiv
3\chi_e u^2+(\chi_\gamma-\chi_e)u
\;\ge\;\tau_\gamma\,,
\label{eq:f_u_PP}
\end{equation}
which determines the allowed region of $u$ (and hence of $x$).

Depending on the signs and relative magnitudes of $(\chi_e,\chi_\gamma)$, Eq.~\eqref{eq:f_u_PP} may admit no solution (stable photon) or define one or two symmetric intervals in $x$. In particular, when $(\chi_\gamma-\chi_e)\le0$ and $\chi_e<0$, one has $f(u)<0$ for all $u\in[0,1/4]$, and photon decay is forbidden.

In the asymptotic regime $\omega^4/(m_e^2\Lambda^2)\gg1$ (i.e., $\tau_\gamma\ll1$), the threshold term $\tau_\gamma$ can be neglected and the condition reduces to
\begin{equation}
u\left[(\chi_\gamma-\chi_e)+3\chi_e u\right]\ge0\,.
\label{eq:PP_asymptotic_condition}
\end{equation}
For $\chi_\gamma\ge\chi_e$ and $\chi_e\ge0$, this inequality is satisfied over the full interval $x\in[0,1]$, reproducing the asymptotic result of Ref.~\cite{Rubtsov:2012kb}. For finite $\tau_\gamma$, however, the inequality may admit no solution below a certain photon energy. The decay becomes kinematically allowed only above a critical value of $\omega$, where a non-empty interval of $x$ satisfying Eq.~\eqref{eq:f_u_PP} first appears. A detailed analysis of the kinematic domains in the $(\chi_e,\chi_\gamma)$ plane is left for a forthcoming work.

\subsection{Amplitude and squared matrix element}
\label{sec:amplitude:decay}

In the case of photon decay, the tree-level amplitude is 
\begin{equation}
\mathcal{A}=-ie\,\epsilon_i(\vec k)\,
\bar{\tilde u}(\vec p_-)\,\gamma^i\,\tilde v(\vec p_+)\,.
\label{eq:PP_amplitude}
\end{equation}

The average squared amplitude is
\begin{equation}
\overline{|\mathcal A|^2}
=\frac{1}{2}\sum_{\text{spins},\,\lambda}|\mathcal A|^2
=\frac{e^2}{2}
\left(\delta_{ij}-\frac{k_i k_j}{\vec k^{\,2}}\right)
\Tr\!\left[\gamma^i(\slashed{\tilde p}_-+m_e)\,
\gamma^j(\slashed{\tilde p}_+-m_e)\right],
\label{eq:PP_A2_trace}
\end{equation}
where the factor $1/2$ now accounts for the average over the two physical photon polarizations.

Evaluating the fermionic trace using standard Dirac algebra yields
\begin{equation}
\Tr\!\left[\gamma^i(\slashed{\tilde p}_-+m_e)\,\gamma^j(\slashed{\tilde p}_+-m_e)\right]
=4\left(
p_-^i p_+^j + p_+^i p_-^j
+\delta^{ij}\,(\tilde p_-\!\cdot\!\tilde p_+ + m_e^2)
\right).
\label{eq:PP_trace}
\end{equation}
Then we have
\begin{equation}
\overline{|\mathcal A|^2}
=4e^2\left[
(\vec p_-\!\cdot\!\vec p_+)
-\frac{(\vec k\!\cdot\!\vec p_-)(\vec k\!\cdot\!\vec p_+)}{\vec k^{\,2}}
+(\tilde p_-\!\cdot\!\tilde p_+)+m_e^2
\right].
\label{eq:PP_A2_compact}
\end{equation}

Using the momentum decomposition \eqref{eq:momentum-decomposition-PP}, the scalar products appearing in Eq.~\eqref{eq:PP_A2_compact} can be expressed in terms of the longitudinal momentum fraction $x$ and the transverse momentum $\vec p_T$. In the ultrarelativistic approximation,
\begin{align}
(\vec p_-\!\cdot\!\vec p_+)+(\tilde p_-\!\cdot\!\tilde p_+)
&\simeq x(1-x)\,\vec k^{\,2}
+\left[\frac{1}{2x(1-x)}-1\right]\vec p_T^{\,2}
+\frac{m_e^2}{2}\,,
\label{eq:PP_scalar1}
\\
\frac{(\vec k\!\cdot\!\vec p_-)(\vec k\!\cdot\!\vec p_+)}{\vec k^{\,2}}
&= x(1-x)\,\vec k^{\,2}\,.
\label{eq:PP_scalar2}
\end{align}

Substituting these expressions into Eq.~\eqref{eq:PP_A2_compact}, the spin-averaged squared amplitude can be written in this case as
\begin{equation}
\overline{|\mathcal A|^2}
\simeq
2e^2\,\frac{[1-2x(1-x)]}{x(1-x)}\,
\left(\vec p_T^{\,2}+m_e^2\right)
+4e^2m_e^2\,.
\label{eq:PP_A2_pT}
\end{equation}

\subsection{Differential decay width}

The decay width for a photon with four-momentum $k^\mu\equiv (\omega,\vec{k})$ in the process $\gamma(k)\to e^-(p_-)\,e^+(p_+)$ is
\begin{equation}
\Gamma_{\gamma\to e^+e^-}
=\frac{1}{2\omega}\int
\frac{d^3\vec p_-}{(2\pi)^3\,2E_-}\,
\frac{d^3\vec p_+}{(2\pi)^3\,2E_+}\,
(2\pi)^4\,
\delta^{(3)}(\vec k-\vec p_--\vec p_+)\,
\delta(\omega-E_--E_+)\,
\overline{|\mathcal A|^2}\,,
\label{eq:Gamma_PP_start}
\end{equation}
where $E_-$ and $E_+$ denote the electron and positron energies, and $\overline{|\mathcal A|^2}$ is computed in Sec.~\ref{sec:amplitude:decay}. 

Parametrizing the final state momenta as in Eq.~\eqref{eq:momentum-decomposition-PP} and proceeding in the same way as with the vacuum Cherenkov emission, we arrive at
\begin{equation}
\Gamma_{\gamma\to e^+e^-}
\simeq \frac{1}{16\pi\,\omega}\int dx\,
\overline{|\mathcal A|^2}\Big|_{\vec p_T^{\,2}=\vec p_T^{\,2}(x)}\equiv \int dx\;\frac{d\Gamma_{\gamma\to e^+e^-}}{dx}\,,
\label{eq:Gamma_PP_simplified_prefactor}
\end{equation}
where $\vec p_T^{\,2}(x)$ is the function given in Eq.~\eqref{eq:pT2_PP}. Using the identity
\begin{equation}
\chi_\gamma-(x^3+(1-x)^3)\chi_e
=(\chi_\gamma-\chi_e)+3x(1-x)\chi_e\,,
\label{eq:PP_identity}
\end{equation}
this function can be written as
\begin{equation}
\vec p_T^{\,2}(x)=
x(1-x)\frac{|\vec k|^4}{\Lambda^2}\Big[(\chi_\gamma-\chi_e)+3x(1-x)\chi_e\Big]-m_e^2\,,
\label{eq:pT2_PP_rate}
\end{equation}
so that the squared amplitude Eq.~\eqref{eq:PP_A2_pT} becomes
\begin{equation}
\overline{|\mathcal A|^2}
\simeq
2e^2\,[1-2x(1-x)]\,
\Big[(\chi_\gamma-\chi_e)+3x(1-x)\chi_e\Big]\,
\frac{|\vec k|^4}{\Lambda^2}
\;+\;4e^2 m_e^2\,.
\label{eq:PP_A2_on_shell}
\end{equation}
Using again $\alpha=e^2/(4\pi)$ and Eq.~\eqref{eq:Gamma_PP_simplified_prefactor}, we find
\begin{equation}
\frac{d\Gamma_{\gamma\to e^+e^-}}{dx}
\simeq
\frac{\alpha\,m_e^2}{\omega}
\left\{
[1-2x(1-x)]\,
\Big[(\chi_\gamma-\chi_e)+3x(1-x)\chi_e\Big]\,
\frac{\omega^4}{2m_e^2\Lambda^2}
\;+\;1
\right\},
\label{eq:Gamma_PP_finalform}
\end{equation}
where $|\vec k|\simeq\omega$ at the order considered.

\subsection{Comparison with the asymptotic limit}

In Ref.~\cite{Rubtsov:2012kb}, the photon-decay width is computed in the asymptotic limit, where the electron mass can be neglected, as an integral over a longitudinal asymmetry parameter $\tilde x\in[-1,1]$ (their Eq.~(17)). Differentiating their expression yields
\begin{equation}
\frac{d\Gamma^\text{asymp}_{\gamma\to e^+e^-}}{d\tilde x}
=
\frac{\alpha}{4}\,
(1+\tilde x^2)\,
\frac{\omega^3}{\Lambda^2}
\left[
\frac{\chi_\gamma}{2}
-\frac{\chi_e}{8}(1+3\tilde x^2)
\right],
\end{equation}
where we have restricted their expression to the case of dimension-six operators and translated their parameters to our notation. Using the change of variables
\begin{equation}
x=\frac{1+\tilde x}{2}\,, 
\qquad
\tilde x=2x-1\,,
\qquad
d\tilde x=2\,dx\,,
\end{equation}
so that
$1+\tilde x^2 = 2[1-2x(1-x)]$
and
$1+3\tilde x^2 = 4[1-3x(1-x)]$,
one finds
\begin{equation}
\frac{d\Gamma^\text{asymp}_{\gamma\to e^+e^-}}{dx}
=
\frac{\alpha\,\omega^3}{2\Lambda^2}\,
[1-2x(1-x)]\,
\Big[(\chi_\gamma-\chi_e)+3\chi_e\,x(1-x)\Big].
\label{eq:Rubtsov_dGdx_x}
\end{equation}

This coincides exactly with Eq.~\eqref{eq:Gamma_PP_finalform} in the limit $m_e\to0$.

\section{Discussion and outlook}
\label{sec:conclusions}

In this work we have developed a consistent dynamical framework for the computation of vacuum Cherenkov radiation and photon decay in a rotationally invariant, parity- and CPT-even, dimension-six Lorentz-invariance-violating extension of QED. Working at leading order in $1/\Lambda^2$, we have derived explicit expressions for the tree-level amplitudes and differential decay widths of the processes $e\to e\gamma$ and $\gamma\to e^+e^-$. Our results extend previous computations~\cite{Rubtsov:2012kb}, which were obtained in the asymptotic ultrarelativistic limit and neglected electron-mass effects. In contrast, the expressions presented here remain valid in the vicinity of the kinematic thresholds, where finite-mass corrections control the opening of the decay channels and are therefore essential for accurate phenomenological applications and Monte Carlo implementations including LIV effects.

A central technical ingredient of our approach is the use of a temporal-derivative representation of the fermionic LIV operator as an alternative to the spatial-derivative implementation employed in Ref.~\cite{Rubtsov:2012kb}. The two formulations agree at the level of physical observables and to leading order in $1/\Lambda^2$. We have shown explicitly that the resulting decay widths reduce to those of Ref.~\cite{Rubtsov:2012kb} in the ultrarelativistic limit $m_e\to 0$. Moreover, the temporal-derivative basis proves technically simpler, since it preserves the standard spatial structure of the fermion--photon vertex and allows one to perform the amplitude calculations using conventional spinor techniques. Although in presenting the modified Lagrangian we have emphasized the generality of our Lorentz-violating QED, this technical simplification suggests that the temporal-derivative basis may be particularly advantageous in future extensions of the framework to more general Lorentz-violating scenarios, including cases in which some of the present assumptions, such as parity or CPT invariance, are relaxed.

Vacuum Cherenkov emission and photon decay constitute the most sensitive probes of Lorentz-violating QED in the high-energy regime. Once kinematically allowed, both channels typically proceed with very large decay rates, so that observational constraints are effectively controlled by the threshold region. This makes a careful characterization of the kinematic domains, including mass corrections, especially relevant. In our analysis this structure is encoded in the dimensionless parameters $\tau_e = m_e^2\Lambda^2/E^4$ and $\tau_\gamma = m_e^2\Lambda^2/\omega^4$, which quantify the relative importance of the mass term with respect to the Lorentz-violating corrections. Depending on the signs and relative magnitudes of the coefficients $(\chi_e,\chi_\gamma)$, either process may be kinematically forbidden or allowed only above a critical energy. Importantly, strong constraints on Lorentz violation from photon or electron instability do not arise exclusively in superluminal scenarios: even when both particles have subluminal dispersion relations, one of the two decay channels may still be open. The structure of the allowed regions therefore reflects the interplay between electron and photon dispersion deformations and makes explicit the correlation between vacuum Cherenkov radiation and photon decay.

The results obtained here provide the theoretical input required for quantitative phenomenological analyses of Lorentz-violating effects in high-energy QED processes. In particular, the explicit expressions for the differential decay widths, valid also near threshold, allow for a controlled implementation in Monte Carlo simulation packages describing the propagation of high-energy particles, the production of secondaries, and the development of electromagnetic cascades in the atmosphere, such as SimProp~\cite{Aloisio:2017iyh}, CRPropa~\cite{CRPropa:2016qdt} or CORSIKA~\cite{Engel:2018akg}, which are widely used in the interpretation of very-high-energy gamma-ray observations. A detailed study of the resulting constraints on $(\chi_e,\chi_\gamma)$, including the combined impact of both decay channels and their implications for astrophysical observations, will be presented in a forthcoming work.

A natural next step is to extend the present dynamical framework to include interaction processes relevant for the production, propagation and detection of high-energy photons and electrons, such as Compton scattering, photon--photon pair production (Breit--Wheeler), Bethe--Heitler pair production and bremsstrahlung. The techniques developed here provide a suitable starting point for such analyses. A systematic study of these interaction channels is currently under investigation, with the goal of achieving a unified and consistently controlled treatment of Lorentz-violating effects in high-energy QED processes.

\acknowledgments{
The authors acknowledge support by the Spanish grant PID2024-160228NB-I00, funded by the Agencia Estatal de Investigación (MCIN/AEI/10.13039/501100011033), and by the Gobierno de Aragón through the research group E21\_23R (Grupo Teórico de Física de Altas Energías). AG is supported by the predoctoral contract CPE\_27\_25 funded by the Gobierno de Aragón under the 2025–2029 programme for predoctoral research personnel. MAR is supported by Project CAPICLE (PDIP-2026-1575) funded by the Croatian Science Foundation (HrZZ). TT and FR acknowledge the support from the University of Rijeka through the project uniri-iz-25-119, and from the Croatian Science Foundation (HrZZ) Project IP-2022-10-4595. JMC, TT and FR acknowledge the support from the Croatian Science Foundation (HrZZ) Project IP-2026-8139. FR gratefully acknowledges the support of the Erasmus Mobility Program of the University of Rijeka. The authors would like to acknowledge the contribution of the COST Actions CA18108 ``Quantum gravity phenomenology in the multi-messenger approach'' and CA23130 ``Bridging high and low energies in search of quantum gravity''.
}
\bibliography{QuGraPhenoBib}

\clearpage
\appendix

\section{Consistency of the collinear approximation}
\label{app:URvalidity}

Here we justify the use of the collinear approximation in the computation of the decay width of vacuum Cherenkov emission when the incoming electron satisfies $|\vec p|\gg m_e$. We use the longitudinal momentum fraction $x$ introduced in Eq.~\eqref{eq:momentum-decomposition},
\begin{equation}
\vec p\,'=(1-x)\vec p-\vec k_T\,,\qquad 
\vec k=x\vec p+\vec k_T\,,\qquad 
\vec k_T\cdot\vec p=0\,.
\label{eq:app:decomposition}
\end{equation}

For $x$ away from the endpoints, $x\in[\delta,1-\delta]$ with $\delta$ sufficiently large, both final particles carry momenta of order $|\vec p|$, and are therefore ultrarelativistic, as the initial particle. In this situation, the leading terms in the energy conservation equation correspond to the massless Lorentz-invariant case, where collinearity is exact. Therefore
\begin{equation}
\frac{\vec k_T^{\,2}}{|\vec p|^2}
=\mathcal{O}\!\left(\frac{m_e^2}{|\vec p|^2},\frac{|\vec p|^2}{\Lambda^2}\right)\ll1,
\end{equation}
and we can use the collinear approximation
Eqs.~\eqref{eq:VC-E-UR}-\eqref{eq:VC-omega-UR}. These equations imply the consistency conditions
\begin{equation}
\frac{m_e^2+k_T^{\,2}}{2(1-x)^2|\vec p|^2}\ll1,\qquad
\frac{\vec k_T^{\,2}}{2x^2|\vec p|^2}\ll1.
\label{eq:coll-conditions}
\end{equation}

Possible violations of this approximation can therefore only arise in the limits $x\to0$ or $x\to1$, where one or more of the conditions in Eq.~\eqref{eq:coll-conditions} may fail.

\paragraph{Limit $x\to0$.}
For a given $|\vec p|$, the second condition in Eq.~\eqref{eq:coll-conditions} would cease to be valid at a value $x_0$ such that
\begin{equation}
x_0^2(|\vec p|)\sim\frac{\vec k_T^2(x_0,|\vec p|)}{|\vec p|^2}.
\label{eq:x_0}
\end{equation}
Here $\vec k_T^2(x,|\vec p|)$ denotes the solution of the energy conservation equation obtained within the collinear approximation,
\begin{equation}
\vec{k}_T^2(x,|\vec p|)
=x^2\left[\frac{|\vec p|^4}{\Lambda^2}\big(x^2(1-x)(\chi_e-\chi_\gamma)+3(1-x)^2\chi_e\big)-m_e^2\right].
\label{eq:kT2_VC_app}
\end{equation}
Although Eq.~\eqref{eq:kT2_VC_app} is derived assuming collinearity, it provides the correct behaviour for $x$ above the point where the approximation could break down, and the limit $x\to x_0$ is continuous.

Substituting Eq.~\eqref{eq:kT2_VC_app} into Eq.~\eqref{eq:x_0} yields
\begin{equation}
1\sim
\left[
\frac{|\vec p|^2}{\Lambda^2}\big(x_0^2(\chi_e-\chi_\gamma)+3\chi_e\big)
-\frac{m_e^2}{|\vec p|^2}
\right].
\label{eq:x0_condition}
\end{equation}
However, the right-hand side is suppressed in the regime considered here, since $|\vec p|^2/\Lambda^2\ll1$ and $m_e^2/|\vec p|^2\ll1$. Therefore Eq.~\eqref{eq:x0_condition} cannot be satisfied, showing that the limit $x\to0$ does not invalidate the collinear approximation. In particular, soft photons do not spoil the consistency of the expansion.

\paragraph{Limit $x\to1$.}
The first condition in Eq.~\eqref{eq:coll-conditions} instead fails when
\begin{equation}
(1-x_1)^2\sim\frac{m_e^2+\vec k_T^2(x_1,|\vec p|)}{|\vec p|^2}.
\label{eq:coll-conditions2}
\end{equation}

Consider now Eq.~\eqref{eq:kT2_VC_app} near $x=1$. We can write
\begin{equation}
m_e^2+\vec{k}_T^{2}(x,|\vec p|)
\approx 2(1-x)m_e^2
+\frac{|\vec p|^4}{\Lambda^2}(1-x)g(x),
\qquad
g(x)=(\chi_e-\chi_\gamma)+3(1-x)\chi_e.
\end{equation}
Condition Eq.~\eqref{eq:coll-conditions2} then implies that any possible breakdown of the collinear approximation is confined to a region satisfying
\begin{equation}
1-x_1=
\mathcal O\left(
\frac{m_e^2}{|\vec p|^2}
+\frac{|\vec p|^2}{\Lambda^2}
\right).
\end{equation}
Since the kinematic condition $\vec k_T^{\,2}\geq 0$ requires $m_e^2/|\vec p|^2 =\mathcal O(|\vec p|^2/\Lambda^2)$ for fixed LIV coefficients of order unity, we obtain
\begin{equation}
1-x_1=
\mathcal O\left(\frac{|\vec p|^2}{\Lambda^2}\right).
\label{eq:x1_momentum}
\end{equation}

We will now examine the impact of the possible loss of validity of the collinear approximation on the decay width. Let $\mathcal I(x)$ denote the integrand of the decay width expressed in terms of the longitudinal fraction $x$,
with $\mathcal I(x)$ understood to vanish outside the
kinematically allowed region.
We can then formally split the width as
\begin{equation}
\Gamma
=\int_0^{x_1}dx\,\mathcal I(x)
+
\int_{x_1}^{1}dx\,\mathcal I(x)
\equiv
\Gamma_{\rm bulk}+\Gamma_{\rm end}.
\end{equation}

The bulk contribution $\Gamma_{\rm bulk}$ lies entirely in the region where the collinear approximation is valid and is therefore computed using the expressions of Sec.~\ref{sec:cherenkov} and \ref{sec:decay}. The endpoint contribution $\Gamma_{\rm end}$ corresponds to an interval of size
\begin{equation}
\Delta x\sim 1-x_1
=\mathcal{O}\!\left(\frac{|\vec p|^2}{\Lambda^2}\right).
\end{equation}

Since the tree-level integrand $\mathcal I(x)$ remains regular near the endpoint, this contribution is bounded by
\begin{equation}
\Gamma_{\rm end}
=\mathcal{O}\!\left(\frac{|\vec p|^2}{\Lambda^2}\right)
\mathcal I_{\rm typ},
\end{equation}
where $\mathcal I_{\rm typ}$ denotes the typical size of the integrand. Within the effective theory domain $|\vec p|\ll\Lambda$, one has $\Gamma_{\rm end}\ll\Gamma_{\rm bulk}$.
We conclude that the decay width is dominated by the region where the collinear approximation holds, while the contribution from the endpoint region near $x\to1$ is suppressed by the small ratio $|\vec p|^2/\Lambda^2$. Since our calculation is performed consistently at this order in the effective expansion, such corrections lie beyond the accuracy of the present result.

An analogous analysis applies to photon decay, $\gamma\to e^-e^+$. The main difference is that, in this case, the collinear approximation may break down near both endpoints, $x\to0$ and $x\to1$, corresponding to one of the final leptons carrying a vanishing fraction of the incoming photon momentum. However, the potentially problematic regions are again restricted to intervals of size $\mathcal O(|\vec k|^2/\Lambda^2)$. Their contributions to the decay width are therefore suppressed by an additional factor of $|\vec k|^2/\Lambda^2$, and do not affect the result at the order considered in our calculation.

This establishes the self-consistency of the collinear kinematic expansions used in Sec.~\ref{sec:cherenkov} and \ref{sec:decay}.

\end{document}